\documentclass{svmult}

\usepackage{mathptmx}
\usepackage{mathptmx}
\usepackage{graphicx}
\usepackage{multicol}
\usepackage{footmisc}
\usepackage{xcolor}
\usepackage{amsmath}
\usepackage{amssymb}
\usepackage{bbold}
\usepackage{float}

\newcommand{\mbi}[1]{\mbox{\boldmath$#1$}}
\newcommand{\bfi}[1]{\mbox{\bf$#1$}}

\newcommand{\mat}[1]{\mbox{\rm\bf #1}}
\newcommand{\lsim}[1]{\mbox{${\,\hbox{\hbox{$ < $}\kern -0.8em \lower 1.0ex\hbox{$\sim$}}\,}$}}
\newcommand{\gsim}[1]{\mbox{${\,\hbox{\hbox{$ > $}\kern -0.8em \lower 1.0ex\hbox{$\sim$}}\,}$}}

\newcommand{\be}{\begin{equation}}
\newcommand{\ee}{\end{equation}}
\newcommand{\ba}{\begin{eqnarray}}
\newcommand{\ea}{\end{eqnarray}}

\def\beqn{\vspace{2mm}
\begin{eqnarray}} 
\def\eeqn{\vspace{2mm} 
\end{eqnarray}}

\begin{document}

\title*{Bayesian analysis of cosmic structures}

\titlerunning{Bayesian analysis of cosmic structures}
\authorrunning{Francisco-Shu Kitaura}


\author{Francisco-Shu Kitaura}
\institute{Karl-Schwarzschild fellow\\
Leibniz-Institut f\"ur Astrophysik Potsdam (AIP), An der Sternwarte 16, D-14482 Potsdam, Germany, \email{kitaura@aip.de}}
\maketitle

\abstract{We revise the Bayesian inference steps required to analyse the cosmological large-scale structure. Here we make special emphasis in the complications which arise due to the non-Gaussian character of the galaxy and matter distribution.  In particular we investigate the advantages and limitations of the Poisson-lognormal model and discuss how to extend this work.
With  the lognormal prior using the Hamiltonian sampling technique and on scales of about 4 $h^{-1}$ Mpc we find that the over-dense regions are excellent reconstructed, however, under-dense regions (void statistics) are quantitatively poorly recovered. Contrary to the maximum a posteriori (MAP) solution which was shown to over-estimate the density in the under-dense regions we obtain lower densities than in N-body simulations. This is due to the fact that the MAP solution is conservative whereas the full posterior yields samples which are consistent with the prior statistics. The lognormal prior is not able to capture the full non-linear regime at scales below $\sim 10$ $h^{-1}$ Mpc for which higher order correlations would be required to describe the matter statistics. However, we confirm as it was recently shown in the context of Ly$\alpha$ forest tomography that the Poisson-lognormal model provides the correct two-point statistics (or power-spectrum).}

\section{Introduction}

   The cosmological large-scale structure encodes a wealth of information about the origin and evolution  of the Universe. A careful study of  the cosmic structure can thus lead to a deeper understanding on structure formation and unveil the cosmological parameters to unprecedented accuracy.
However, the data are plagued by many observational effects like the mask and selection function of the particular surveys and the bias related to the matter tracer ({e.~g.~}galaxies). It is thus clear that a statistical treatment is necessary to compare observations with theory and perform a detailed study of structure formation.

In this { report} we focus on the systematic effects which arise from observed data and present in detail the Bayesian approach (Section \ref{sec:bayes}) to describe the statistics of the data and the large-scale structure.  

Finally, we show some numerical experiments which unveil the state-of-the-art in the field and the problems which should be addressed in future work.

\section{Bayesian approach}
\label{sec:bayes}


The evolution from a Gaussian homogeneous Universe to a complex non-linear and non-Gaussian cosmic web can be accurately modeled  with $N$-body simulations (see e.~g.~\cite{2005Natur.435..629S}). 
Hence we can test the different statistical models describing the nature of the matter distribution simplifying our model selection process. 

In this context a Bayesian approach is ideal as it clearly incorporates the assumptions in form of { conditional} probability distribution functions (PDFs) making a distinction between the model for the observed/measured data ${\bf d}$ represented by the likelihood and the model for the seeked signal ${\bf s}$ represented by the prior (${ P}({\bf d}|{\bf s},{\bf p})$ and ${ P}({\bf s}|{\bf p})$, respectively).
Note that we have to condition all the PDFs to some set of parameters ${\bf p}$
which encode our prior knowledge.
Bayes theorem yields the posterior:
${ P}({\bf s}|{\bf d},{\bf p})=\frac{{ P}({\bf s}|{\bf p}){ P}({\bf d}|{\bf s},{\bf p})}{\int {\rm d}{\bf s}\,{ P}({\bf s}|{\bf p}){ P}({\bf d}|{\bf s},{\bf p})}$. In short: Posterior=prior$\times$likelihood/evidence. The evidence which can be important for model comparison and selection can be simply considered as a  normalization constant for our purposes.

\subsection{Bayesian inference steps}  

From Bayes theorem we can already extract the necessary ingredients to perform a Bayesian analysis. First the prior and the likelihood have to be { defined} to find an expression for the posterior PDF.  From the posterior one may obtain an estimate of the signal either computing the maximum or sampling the full posterior PDF. Here we enumerate the different steps:
  \begin{enumerate}
  \item Definition of the prior: knowledge of the underlying signal  
  \item Definition of the likelihood: nature of the observed data
  \item Linking the prior to the likelihood: link between signal and data 
  \item Bayes theorem: definition of the posterior
  \item Maximization of the posterior: MAP
  \item Sampling the posterior: MCMC
  \end{enumerate}

\subsection{Definition of the prior: knowledge of the underlying signal}

The prior distribution function describes the statistical nature of the signal we want to recover from degraded measured data. In our case we want to obtain the three dimensional map of the large-scale structure represented by the matter over-density field ${\bf \delta_{\rm M}}$. { For computational reasons we choose an equidistant grid with $N_{\rm c}$ cells which permits us to use fast Fourier transforms.}

\subsubsection{Gaussian prior: cosmic variance and cosmological parameters}

The most simple PDF to characterize a cosmic field with a given power-spectrum is represented by the Gaussian distribution \cite{1986ApJ...304...15B}:
\begin{equation}
{ P}({\bf \delta_{\rm M}}|{\bf p})= \frac{1}{\sqrt{(2\pi)^{N_{\rm c}}\det({\mat S}_\delta)}}\exp\left[-\frac{1}{2}{\bf \delta_{\rm M}}^\dagger{\mat S}_\delta^{-1}{\bf \delta_{\rm M}}\right]{,}
\end{equation} 
with ${\bf p}$ being the set of cosmological parameters which determine the auto-correlation matrix { ${\mat S}_\delta\equiv\langle{\bf \delta_{\rm M}}^\dagger{\bf \delta_{\rm M}}\rangle$} or its Fourier transform, the power-spectrum { $P_\delta(\bf k)$ (where $\bf k$ is the $k$-vector in Fourier space)}.
We know however that the matter statistics is skewed due to gravitation. We need non-Gaussian models to better characterize the matter field.

\subsubsection{Non-Gaussian priors}

 The Gaussian distribution function can be expanded with the Edgeworth expansion \cite{1995ApJ...442...39J}. However, this is only valid for moderate non-Gaussian fields. 
One can instead make a variable transformation of the Gaussian variable and apply the lognormal assumption \cite{1991MNRAS.248....1C}.
Such a distribution function may also be expanded leading to very accurate fits in the univariate matter statistics comparing to $N$-body simulations  \cite{colombi}.

Let us introduce the  field ${\bfi\Phi}$ which has zero mean by definition { for each cell $i$}: 
   \begin{equation}
     \Phi_i\equiv \ln \rho_i- \langle\ln \rho \rangle=\ln(1+\delta_{{\rm M}i})-\mu_i \,.
  \end{equation}  
Then the multivariate Edgeworth expansion is given by \cite{kitaura_skewed}:
 \begin{eqnarray}	
\lefteqn{P({\bfi\Phi})= { G({\bfi\Phi})}\left[1+\frac{1}{3!}\sum_{i'j'k'}\langle\Phi_{i'}\Phi_{j'}\Phi_{k'}\rangle_{\rm c}\sum_{ijk}S_{ii'}^{-1/2}S_{jj'}^{-1/2}S_{kk'}^{-1/2} h_{ijk}{({\mat S^{-1/2} \bfi \Phi})}\right.}\nonumber\\
&&\hspace{-1cm}\left.+\frac{1}{4!}\sum_{i'j'k'l'}\langle\Phi_{i'}\Phi_{j'}\Phi_{k'}\Phi_{l'}\rangle_{\rm c}\sum_{ijkl}S_{ii'}^{-1/2}S_{jj'}^{-1/2}S_{kk'}^{-1/2}S_{ll'}^{-1/2} h_{ijkl}{({\mat S^{-1/2} \bfi \Phi})}+\dots\right]\,, 	  
 \end{eqnarray} 
  with $G({\bfi\Phi})$ being a Gaussian PDF with zero mean and variance  { ${\mat S}\equiv\langle{\bfi \Phi}^\dagger{\bfi \Phi}\rangle$}   for the variable ${\bfi\Phi}$, $\langle\Phi_{i'}\Phi_{j'}\Phi_{k'}\rangle_{\rm c}$ and $\langle\Phi_{i'}\Phi_{j'}\Phi_{k'}\Phi_{l'}\rangle_{\rm c}$ the third and fourth order cumulants, and with $h_{ijk}$ and  $h_{ijkl}$ being the third and fourth order Hermite polynomials.

\subsubsection{Lognormal model}   

 The multivariate lognormal model 	(${ { P}({\bf\delta_{{\rm M}}}|{\mat S})=G(\bfi \Phi})$) is given by \cite{kitaura_log}:
 \begin{eqnarray}
	\lefteqn{{ P}({\bf\delta_{{\rm M}}}|{\mat S})=\frac{1}{\sqrt{(2\pi)^{N_{\rm c}} {\rm det} ({\mat S})}}\prod_k 		\frac{1}{1+\delta_{{\rm M}k}}} \\
	&&\hspace{-0.5cm}\times {\rm exp}\left({-\frac{1}{2}\sum_{ij} \left({\rm ln}(1+\delta_{{\rm M}i})-\mu_i\right) 					S^{-1}_{ij} \left({\rm ln}(1+\delta_{{\rm M}j})-\mu_j\right)}\right) \, {.}\nonumber
	\end{eqnarray}
Note that this PDF converges to the Gauss distribution when { $|\delta_{\rm M}|\ll 1$}. 

\subsection{Definition of the likelihood: nature of the observable}  

A galaxy sample represents a discrete biased sample of the underlying matter field. Its distribution can be sub- or super-Poisson depending on local and non-local properties \cite{1996MNRAS.282..347M,2001MNRAS.320..289S,2002MNRAS.333..730C}.
Based on  a  discrete version of the Press-Schechter formalism \cite{1983MNRAS.205..207E} found a  Borel distribution.  
Another non-Poisson distribution was found in the context of a thermodynamical description of gravity \cite{1984ApJ...276...13S,1988ApJ...331...45I}.
In \cite{1995MNRAS.274..213S} it is shown that both distribution functions can be identical under certain assumptions.
Let us write the gravitothermal dynamics distribution function generalized to have a scale dependent parameter ${\mat Q}$:
  \begin{eqnarray}
      \label{eq:Poissonian}
      \lefteqn{{ P}({{\mbi N}}|{{\mbi \lambda}}, {\mat Q}) =\prod_k \frac{\sum_{j}(\delta^{\rm K}_{k,j}-Q_{k,j})\lambda_j}{{N_k}!}}\\
      &&\times {\left(\sum_{l}(\delta^{\rm K}_{k,l}-Q_{k,l})\lambda_l+\sum_{m}Q_{k,m}N_m\right)}^{N_k-1} {\rm exp}\left({-\sum_{n}(\delta^{\rm K}_{k,n}-Q_{k,n})\lambda_n-\sum_{o}Q_{k,o}N_o}\right) \nonumber\,
    \end{eqnarray}
Note that this PDF { simplifies to} the Poisson { distribution} when ${\mat Q}$ is zero.

  \subsubsection{Poisson limit}
For a sparse sample we can assume a Poisson distribution \cite{Peebles-80}:   
    \begin{equation}
      \label{eq:Poisson}
            { P}({\bf N}|{\bf\lambda}) =\prod_k \frac{\lambda_k^{N_k}{\rm exp}\left({-\lambda_k}\right)}{{N_k}!}\,.
    \end{equation}
Note that usually only the Poisson variance has been considered in the context of large-scale structure reconstructions \cite{zaroubi,WienerFSL,2004MNRAS.352..939E}. The full treatment was introduced by \cite{kitaura,kitaura_log} and applied to the Sloan Digital Sky Survey \cite{jasche_sdss}.

\subsection{Link between the prior and the likelihood}   
  
The link between the observed/measured data and the signal is usually not trivial and needs to be modeled to find the posterior distribution function.  
In particular we seek a relation between the expected number counts $\lambda$ and the  signal we want to recover ${\bf \delta_{{\rm M}}}$
In our case we have three main complications: the galaxy bias, the completeness of the survey and the  uncertainties in the redshift positions.

\subsubsection{Galaxy bias}
   
The relation between the galaxy ${\bf \delta_{\rm g}}$ and matter ${\bf \delta_{{\rm M}}}$ density fields is non-local and non-linear \cite{Cooray-Sheth-02}.
Let us write such a general relation as:
	\begin{equation}
	\delta_{{\rm g}i}=B({\bf\delta_{{\rm M}}})_i  {,}
	\end{equation} 
One may parametrize this relation expanding the density field { as in} \cite{1993ApJ...413..447F}:  
	\begin{equation}
	\delta_{{\rm g}i}=\sum_j B^1_{ij}\delta_{{\rm M}j}+\delta_{{\rm M}i}\sum_j B^2_{ij}\delta_{{\rm M}j}+\dots  {,}
	\end{equation}
here generalized to be non-local with the scale dependent bias parameters $B^1_{ij}$, $B^2_{ij}$, $\dots$.
Non-local transformations of the density field should be further investigated.
{  Here one may incorporate the halo model into the Bayesian framework (see the recent works on halo model based reconstructions by \cite{2009ApJ...702..249R,2009MNRAS.394..398W}).}

\subsubsection{Response operator}   

The response operator ${\mat R}$ should encode the sky mask, radial selection function and may even encode the uncertainty in the redshift position of galaxies.
 In general such a relation is not trivial:
\begin{equation}
\label{eq:data_model}
\lambda_i= \lambda_i({\bf\delta_{{\rm M}}})=R({\bf\delta_{{\rm g}}}({\bf\delta_{{\rm M}}}))_i\, .
\end{equation}
If we focus our attention to the completeness ${\bf w}$ then we can write:
\begin{equation}
\label{eq:data_model}
\lambda_i=w_i \bar{N}(1+B({\bf \delta_{{\rm M}}})_i)\, ,
\end{equation}
with $\bar{N}$ being the mean number of galaxies in the observed volume.
Assuming a linear bias relation $b$ this expression is reduced to:
\begin{equation}
\label{eq:data_model}
\lambda_i=w_i \bar{N}(1+b\delta_{{\rm M}i})\, .
\end{equation}

 \subsection{Bayes theorem: the posterior}   
Armed with the prior, the likelihood and the link between both we can apply Bayes theorem to obtain the posterior PDF.
A general expression for such a posterior PDF { can be obtained by} plugging in what we have discussed in previous sections (an expanded lognormal prior and a non-Poissonian likelihood):
\begin{eqnarray}
\lefteqn{{ P}({\bf\delta_{{\rm M}}}|{\bf N},{\mat S})}\nonumber\\
&&\hspace{0.cm}\propto \left\{\prod_l \frac{1}{1+\delta_{{\rm M}l}} {\rm exp}\left({-\frac{1}{2}\sum_{ij} \left({\rm ln}\left(1+\delta_{{\rm M}i}\right)-\mu_i\right) S^{-1}_{ij} \left({\rm ln}\left(1+\delta_{{\rm M}j}\right)-\mu_j\right)}\right)\right\}\nonumber\\
&&\hspace{0.cm}\times\left\{[1+\frac{1}{3!}\sum_{i'j'k'}\langle\Phi_{i'}\Phi_{j'}\Phi_{k'}\rangle_{\rm c}\sum_{ijk}S_{ii'}^{-1/2}S_{jj'}^{-1/2}S_{kk'}^{-1/2} h_{ijk}{({\mat S^{-1/2} \bfi \Phi})}\right.\nonumber\\
&&\hspace{0.cm}\left.+\frac{1}{4!}\sum_{i'j'k'l'}\langle\Phi_{i'}\Phi_{j'}\Phi_{k'}\Phi_{l'}\rangle_{\rm c}\sum_{ijkl}S_{ii'}^{-1/2}S_{jj'}^{-1/2}S_{kk'}^{-1/2}S_{ll'}^{-1/2} h_{ijkl}{({\mat S^{-1/2} \bfi \Phi})}+\dots]\right\}\nonumber\\
&& \hspace{0.cm}\times\left\{ \prod_k \frac{\sum_{j}(\delta^{\rm K}_{k,j}-Q_{k,j})w_j \bar{N}(1+B({\bf\delta_{{\rm M}}})_j)}{{N^{\rm g}_k}!} \right.\nonumber\\
&& \hspace{0.cm}\times{\left(\sum_{l}(\delta^{\rm K}_{k,l}-Q_{k,l})w_l \bar{N}(1+B({\bf\delta_{{\rm M}}})_l)+\sum_{m}Q_{k,m}N_m^{\rm g}\right)}^{N^{\rm g}_k-1}\nonumber\\
      &&\hspace{0.cm}\times \left.{\rm exp}\left({-\sum_{n}(\delta^{\rm K}_{k,n}-Q_{k,n})w_n \bar{N}(1+B({\bf\delta_{{\rm M}}})_n)-\sum_{o}Q_{k,o}N_o^{\rm g}}\right)\right\} {,}
\end{eqnarray}
If we assume a lognormal prior, a Poisson likelihood and a linear bias relation we get \cite{kitaura_log,kitaura_lyman}
:
{ 
\ba
\lefteqn{{ P}({\bfi\Phi}|{\bf N},{\mat S})\propto G(\bfi\Phi)}\\
&&\hspace{-.5cm}\times\prod_k\frac{\left(w_k \bar{N}\left(1+b\left(\exp\left(\Phi_k+\mu\right)-1\right)\right)\right)^{N_k}{\rm exp}\left({-w_k \bar{N}\left(1+b\left(\exp\left(\Phi_k+\mu\right)-1\right)\right)}\right)}{{N_k}!}\nonumber\,,
\ea
where we have used the lognormal transformation relating the nonlinear density field $\delta_{\rm M}$ to its Gaussian component $\bfi\Phi$ through $\delta_{{\rm M}i}=\exp\left(\Phi_i+\mu\right)-1$.
}

\subsection{Maximum a posteriori}

Once we have an analytical expression for the posterior distribution function we can compute the maximum of that distribution (MAP).  
The MAP solution for the signal ${\bf s}$ is obtained by searching for the extrema of the {\it energy} { $E({\bf s})\equiv -\ln\left({ P}\left({\bf s}|{\bf d},{\bf p}\right)\right)$}:
	\begin{equation}
	\frac{\partial E({\bf s})}{\partial s_l}=0  {,}
	\end{equation}
Here efficient schemes are crucial to deal with large cell numbers on which the density has to be computed. 
Iterative schemes have been shown to cope with this problem \cite{kitaura,kitaura_log}.

 \subsection{Sampling the posterior}   
   
Alternatively one may want to sample the full posterior distribution. { Until now we have assumed that the power-spectrum is known and that the data have been previously converted to real-space correcting for redshift distortions. However, it is desirable to consistently estimate the peculiar velocity field ${\bf v}$ and relax the dependence on the cosmological model by jointly sampling the power-spectrum.
  This can be done splitting the full problem into simpler ones with conditional probability distribution functions. In particular with the Gibbs-sampling scheme  one can sample from the joint PDF ${P({\bfi \delta_{\rm M}},  {\bf v}, \mat S|{\bf d^z})}$
  of the matter density field ${\bfi \delta_{\rm M}}$, the peculiar velocity field ${\bf v}$ and the covariance (or power-spectrum) $\mat S$  given some nonlinear data in redshift space ${\bf d}^z$ as follows
\ba
\hspace{2cm}{\bfi \Phi}^{(j+1)}&\hookleftarrow& P({\bfi \Phi}\mid{\bf v}^{(j)},\mat S,{\bf d^z}){,}\label{eq:sig} \\
\hspace{2cm}{\mat S^{(j+1)}}&\hookleftarrow& P({\mat S}\mid{\bfi \Phi}^{(j+1)}) \label{eq:pow}{,}\\
\hspace{2cm}{\bf v}^{(j+1)}&\hookleftarrow& P({\bf v}\mid{\bfi \Phi}^{(j+1)}) \label{eq:vel}{,}
\ea
with the arrows standing for the corresponding sampling process \cite{gibbsamp,toolsstatinf,2004ApJ...609....1J,2004PhRvD..70h3511W,2004ApJS..155..227E,kitaura,jasche_gibbs,kitaura_lyman}.

First, the matter density field (Eq.~\ref{eq:sig}) can be sampled with the Hamiltonian sampling scheme  \cite{duane,2008MNRAS.389.1284T,jasche_hamil,kitaura_lyman} under the Gaussian prior assumption for the variable ${\bfi \Phi}$ and encoding the lognormal transformation between the linear and the nonlinear density fields ($\delta_{{\rm M}i}=\exp\left(\Phi_i+\mu\right)-1$) in the likelihood \cite{viel,2009ApJ...698L..90N,2011ApJ...731..116N}.
Second, the power-spectrum corresponding to ${\bfi \Phi}$ (Eq.~\ref{eq:pow}) can be consistently sampled with the inverse Gamma distribution function \cite{kitaura,jasche_gibbs}.
Finally, the peculiar velocity sampling (Eq.~\ref{eq:vel})  which permits us to do the mapping between real- and redshift-space can be done with Lagrangian perturbation theory from the Gaussian component of the density field \cite{1995A&A...298..643H,1999MNRAS.308..763M,kitaura,kitaura_lyman}. 
}

\section{Numerical experiments}
\label{sec:num}

Here we demonstrate the numerical computation of the multivariate non-Gaussian matter field statistics and its power-spectrum.  We will restrict ourselves in
this work to the lognormal prior and the Poisson likelihood. {  It was shown in \cite{kitaura_lyman} how to sample the power-spectrum consistently with this model. However, we will show here that even with a fix prior for the power-spectrum one can extract the underlying features and the correct shape of the power-spectrum since the dependence on the prior becomes sub-dominant in the presence of {\it good enough} data as it is provided with present galaxy redshift surveys.}

\subsection{Setup}

 We construct the mock  observed data taking a random sub-sample of the particles in  the Millennium run at redshift zero \cite{2005Natur.435..629S} which was gridded on a $128^3$ mesh.   Our setup permits us to avoid the biasing problem in our tests. Note that we also avoid the redshift distortions by considering the dark matter particles in real-space.
The mocks were generated with a radial selection function using an exponential decaying model of completeness $w$  \cite{kitaura_sdss}. The final mock galaxy samples have 350961 particles. The observer was set at the center of the box, i.e.~at coordinates: X=250 $h^{-1}$ Mpc, Y=250 $h^{-1}$ Mpc, and Z=250 $h^{-1}$ Mpc.
  We calculate the power-spectrum $P_\delta({\bf k})$ which determines the covariance matrix ${\mat S}$ with a
 nonlinear fit which also describes the effects of { virialized}
 structures including a halo term as given by
 \cite{2003MNRAS.341.1311S} at redshift $z=0$.  
We apply the Hamiltonian scheme  with the \textsc{ARGO}-code \cite{kitaura,kitaura_log,kitaura_lyman} to sample the full posterior distribution function.  

\subsection{Results}

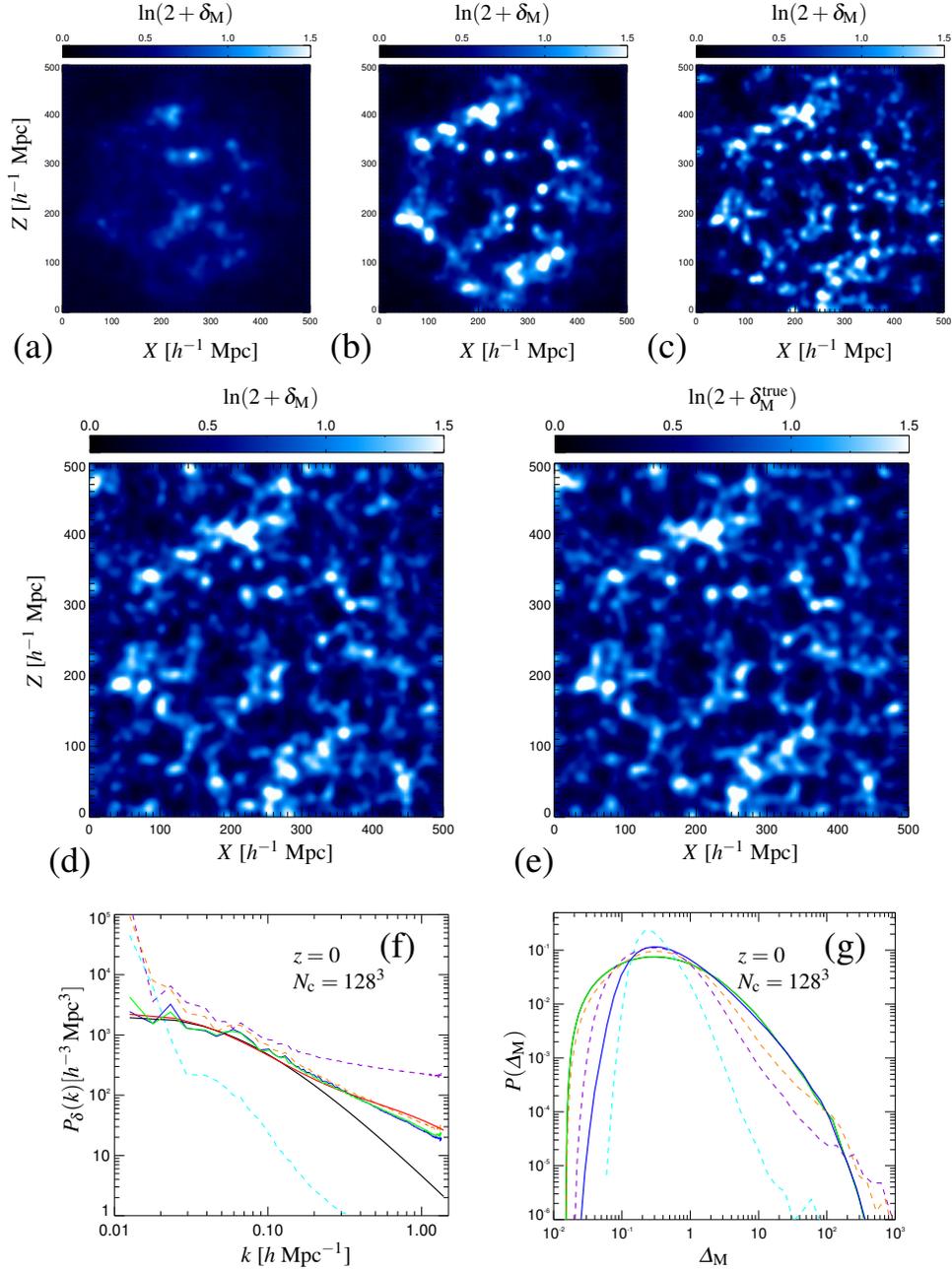
\begin{figure*}
\begin{tabular}{cc}
\hspace{-0.5cm}
\includegraphics[width=4.2cm]{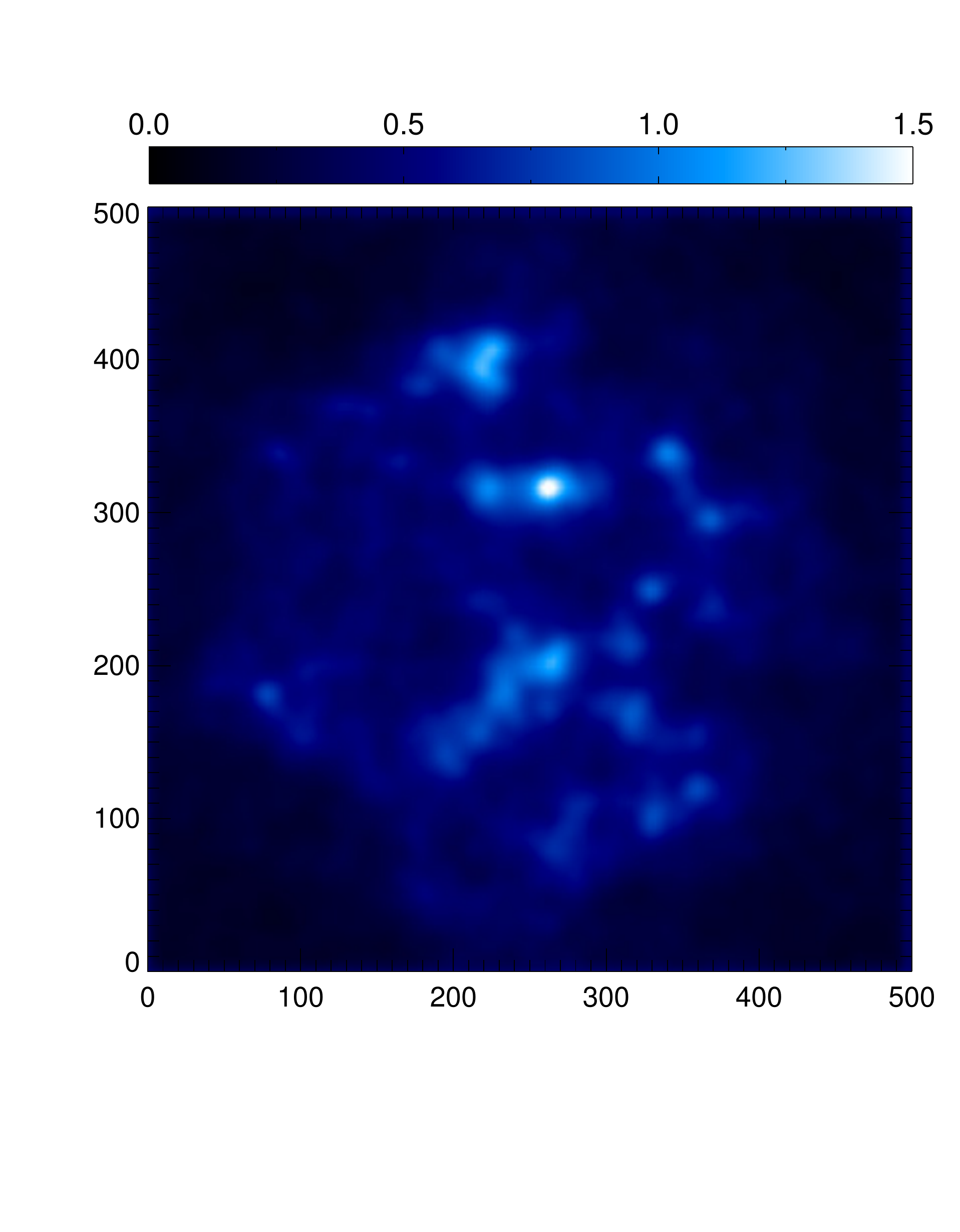}
\put(-122,83){\rotatebox[]{90}{{$Z$ [$h^{-1}$ Mpc]}}}
\put(-70,15){{$X$ [$h^{-1}$ Mpc]}}
\put(-72,145){{$\ln(2+\delta_{\rm M})$}}
\put(-120,15){{\Large (a)}}
\hspace{0.cm}
\includegraphics[width=4.2cm]{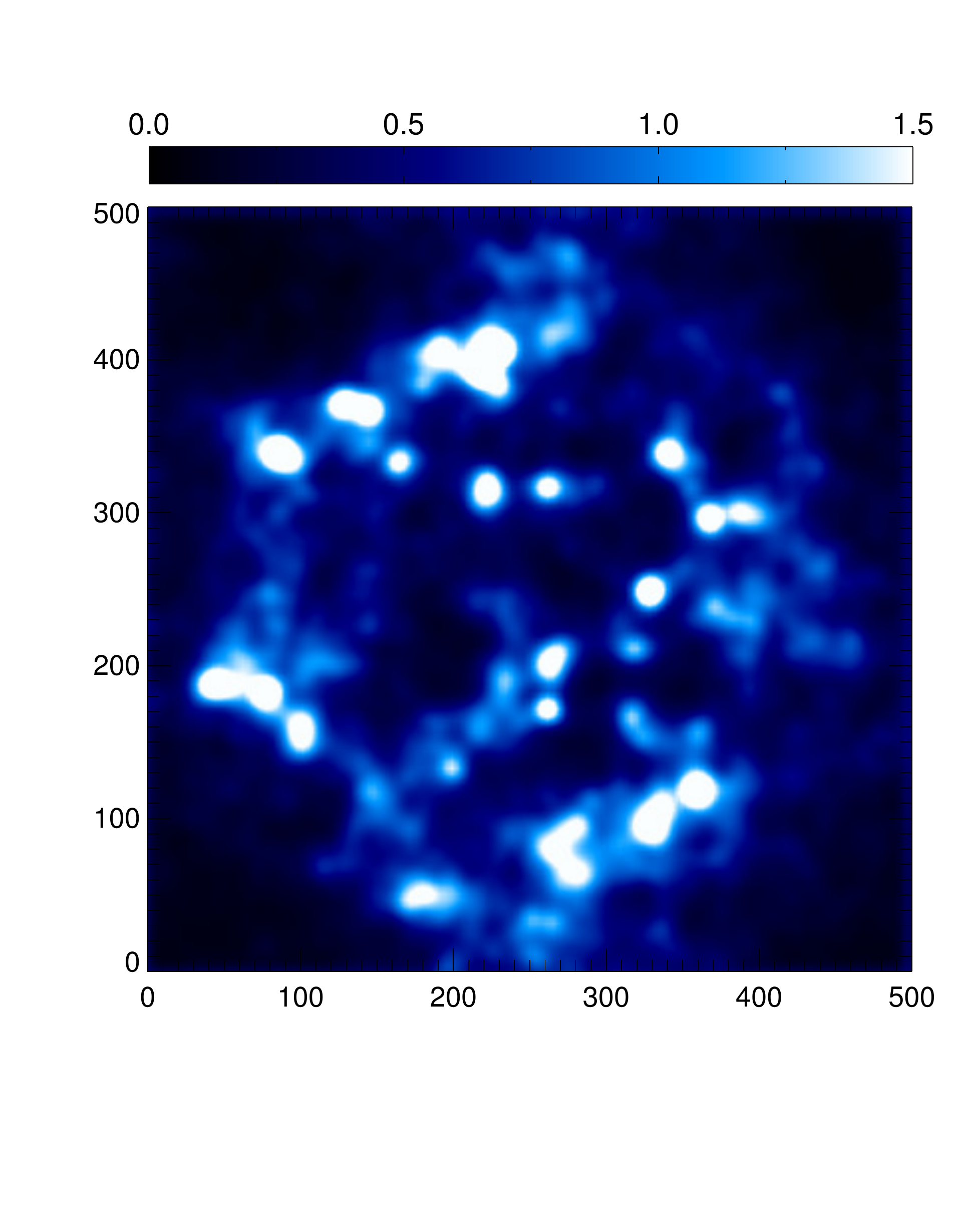}
\put(-70,15){{$X$ [$h^{-1}$ Mpc]}}
\put(-72,145){{$\ln(2+\delta_{\rm M})$}}
\put(-120,15){{\Large (b)}}
\hspace{0.cm}
\includegraphics[width=4.2cm]{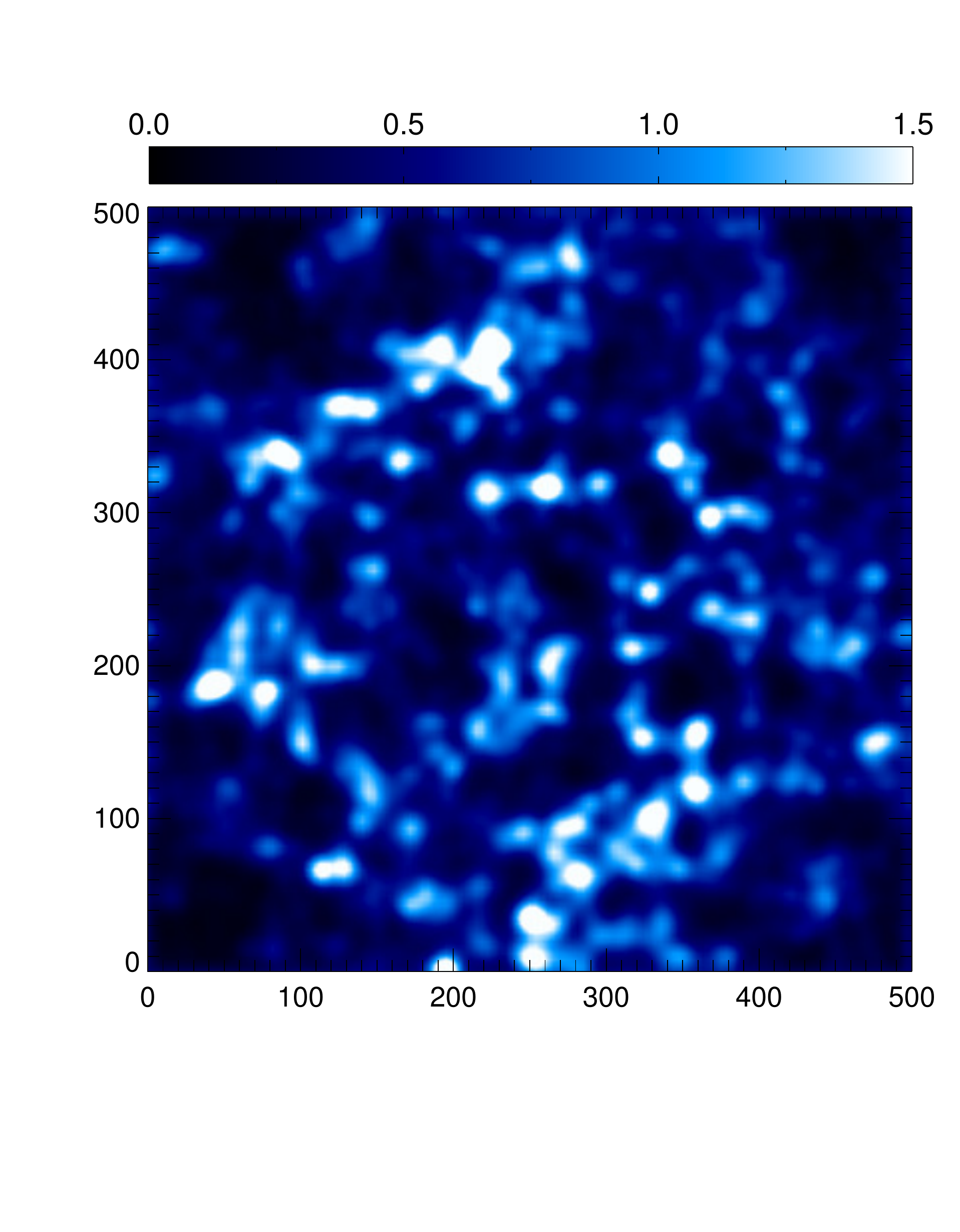}
\put(-70,15){{$X$ [$h^{-1}$ Mpc]}}
\put(-72,145){{$\ln(2+\delta_{\rm M})$}}
\put(-120,15){{\Large (c)}}
\vspace{-0.5cm}
\\
\hspace{-0.8cm}
\includegraphics[width=6cm]{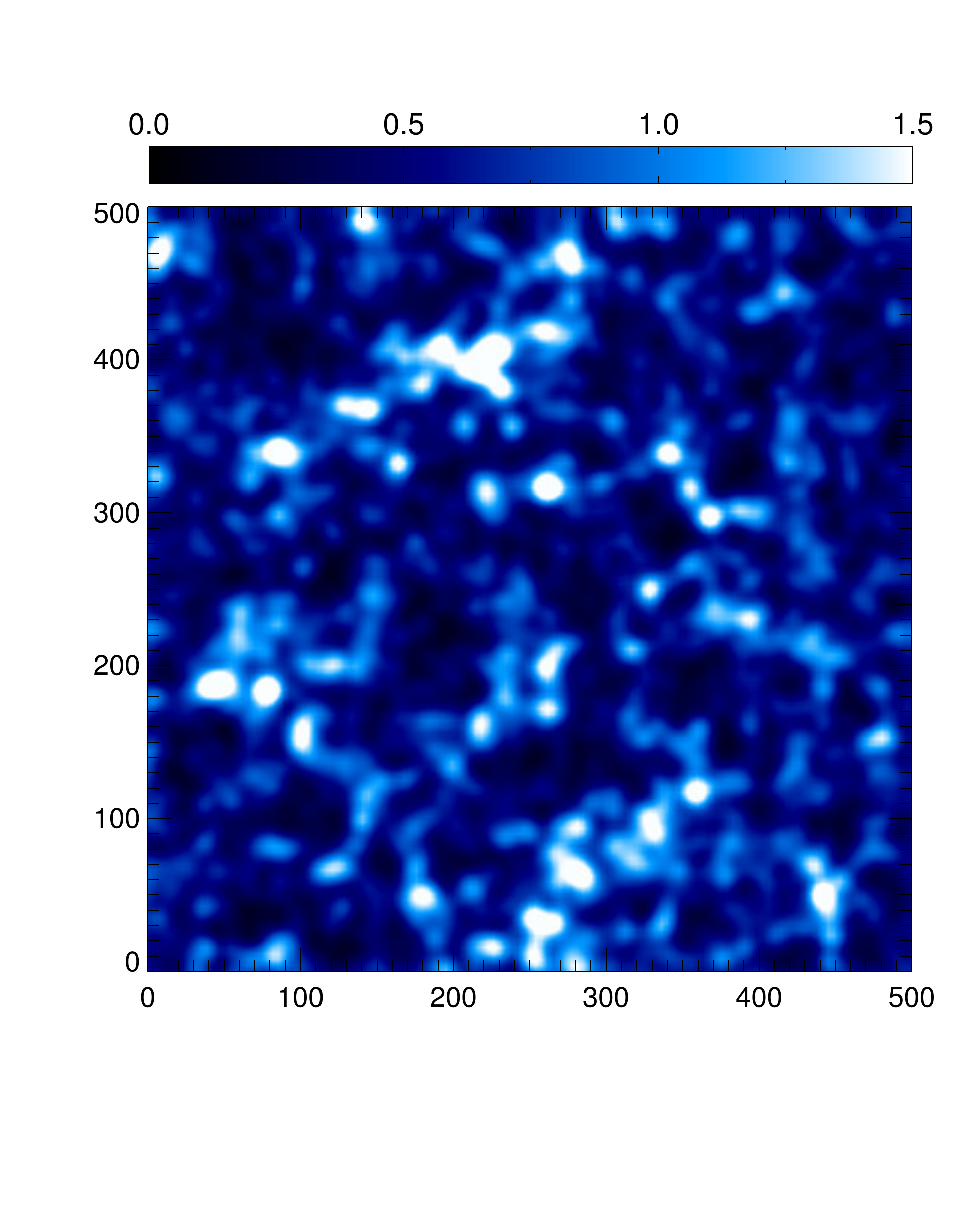}
\put(-170,115){\rotatebox[]{90}{{$Z$ [$h^{-1}$ Mpc]}}}
\put(-95,30){{$X$ [$h^{-1}$ Mpc]}}
\put(-93,205){{$\ln(2+\delta_{\rm M})$}}
\put(-160,25){{\Large (d)}}
\hspace{0.2cm}
\includegraphics[width=6cm]{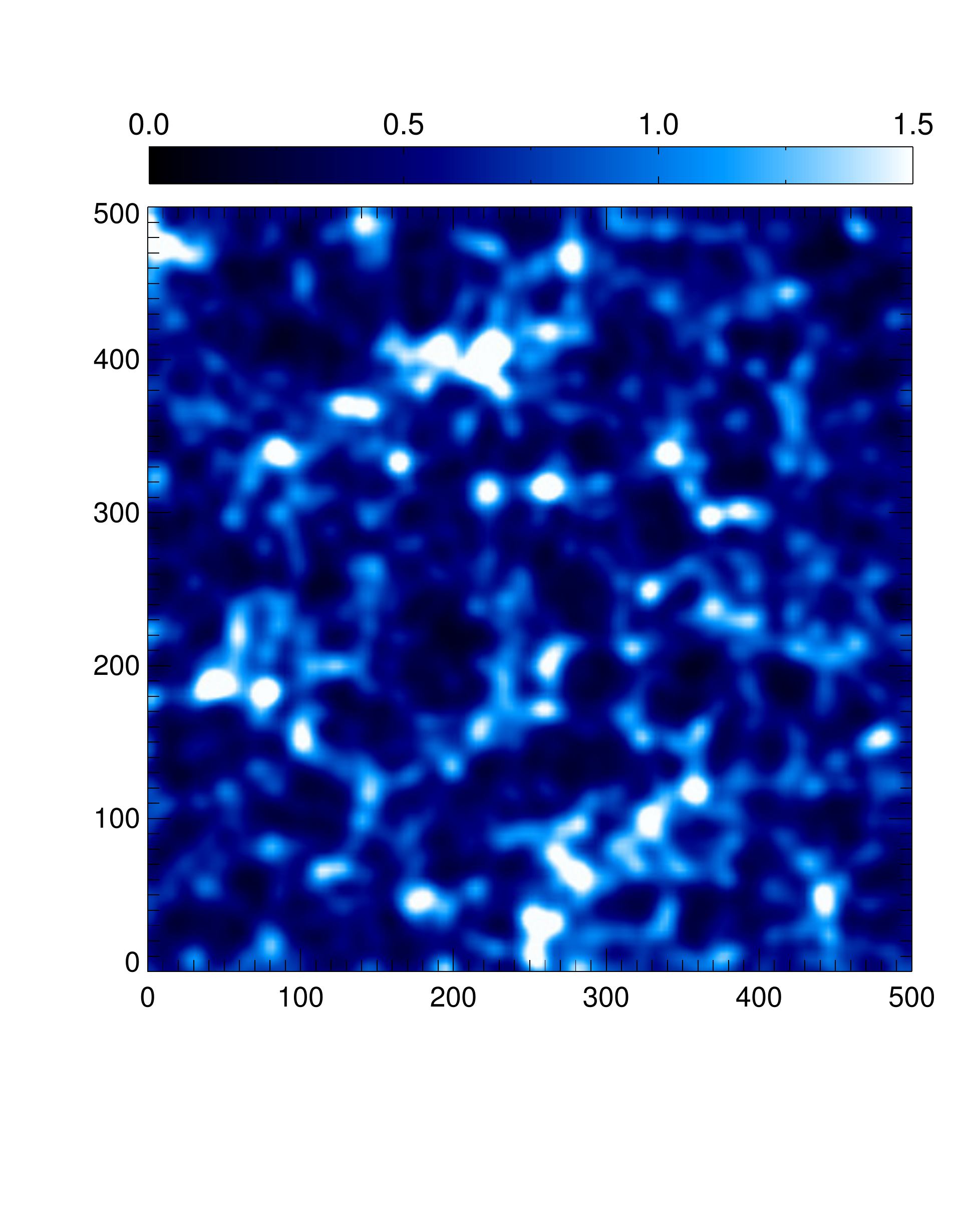}
\put(-95,30){{$X$ [$h^{-1}$ Mpc]}}
\put(-93,205){{$\ln(2+\delta^{\rm true}_{\rm M})$}}
\put(-160,25){{\Large (e)}}
\vspace{-.8cm}
\\
\hspace{-0.8cm}
\includegraphics[width=6.cm]{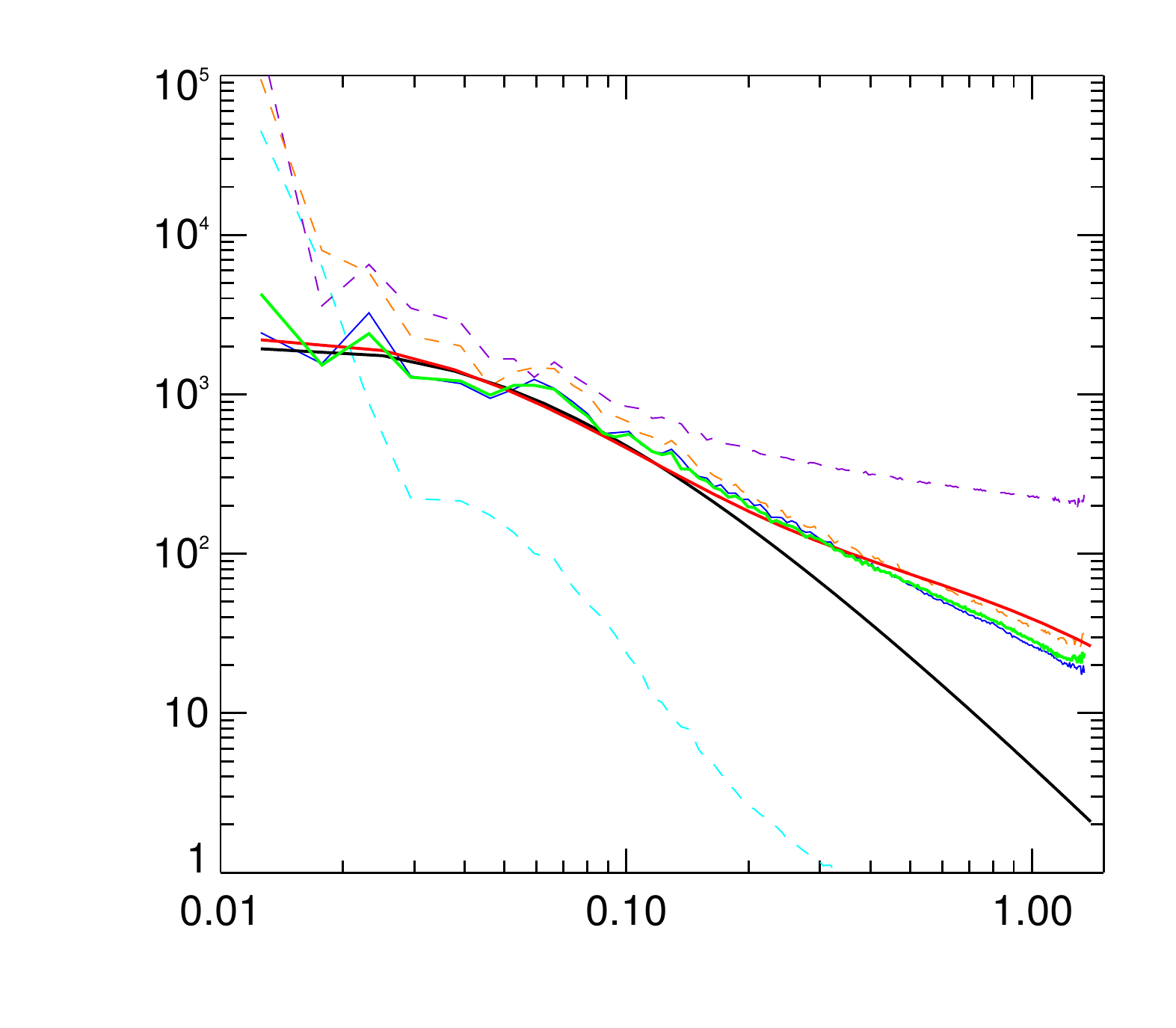}
\put(-160,80){\rotatebox[]{90}{{$P_\delta(k) [h^{-3}\,{\rm Mpc}^3]$}}}
\put(-90,5){{$k$ [$h$ Mpc$^{-1}$]}}
\put(-70,120){{$z=0$}}
\put(-70,110){{$N_{\rm c}=128^3$}}
\put(-37,122){{\Large (f)}}
\includegraphics[width=6.cm]{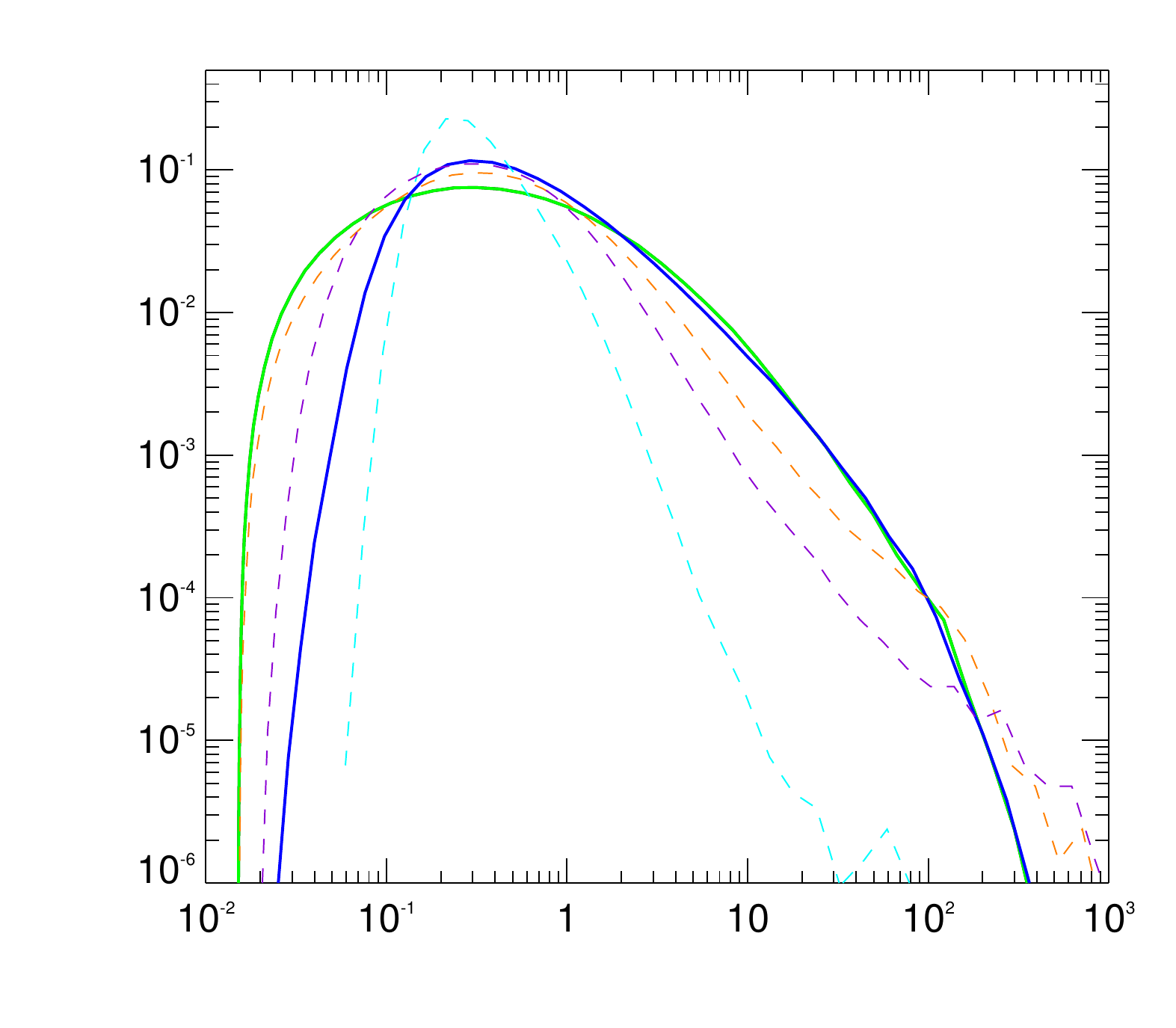}
\put(-160,80){\rotatebox[]{90}{{$P(\Delta_{\rm M})$}}}
\put(-85,5){{$\Delta_{\rm M}$}}
\put(-70,120){{$z=0$}}
\put(-70,110){{$N_{\rm c}=128^3$}}
\put(-37,122){{\Large (g)}}
\vspace{-.0cm}
\end{tabular}
\caption{ \label{fig:results} Panels { (a), (b), (c), (d), (e)}: Mean over 10 neighboring slices around the center of a 500 [$h^{-1}$ Mpc] box with $128^3$ cells based on the Millennium Run after Gaussian convolution with smoothing radius 10 $h^{-1}$ Mpc (cell resolution $\sim3.9$ [$h^{-1}$ Mpc]). Panel { (a)}: Reconstruction of a Poisson distributed mock point-source sample including radial selection effects corresponding to $w_1$ in \cite{kitaura_log} (about $3.5\times10 ^{5}$ particles) after 1 iteration. Same as previous panel after: Panel { (b)}:  3 iterations, Panel { (c)}:  10 iterations and
Panel { (d)}:  2000 iterations.  Panel { (e)}: complete sample (about $10^{10}$ matter tracers).  Panel { (f)}: Blue curve: measured power spectrum  of the Millennium Run. Black curve: linear power spectrum.   Red curve: assumed nonlinear power spectrum. Cyan dashed curve: 1st sample. Purple curve: 3rd sample. Orange dashed curve: 10th sample. Green curve: sample 2000.  Panel { (g)}:  Blue continuous curve: measured matter statistics of the Millennium Run after griding the dark matter particles and binning the $\Delta_{\rm M}\equiv 1+\delta_{\rm M}$ over-density field with 0.03 spacings.  The rest of the curves correspond to the samples with the same color coding as in the previous panel.} 
\label{fig:rec}
\end{figure*}

Our results show the evolution of the density samples as the number of iterations increases together with its corresponding matter statistics and power-spectra (see Fig.~\ref{fig:results}).

  We find that on scales of about 4 $h^{-1}$ Mpc the over-dense regions are excellent reconstructed, however, under-dense regions (void statistics) are quantitatively poorly recovered (compare dark blue and green curves in panel g). Contrary to the maximum a posteriori (MAP) solution which was shown to over-estimate the density in the under-dense regions \cite{kitaura_log} we obtain lower densities than in $N$-body simulations. This is due to the fact that the MAP solution is conservative whereas the full posterior yields samples which are consistent with the prior statistics.   The lognormal prior is not able to capture the full non-linear regime at scales below $\sim 10$ $h^{-1}$ Mpc for which higher order correlations would be required to describe the matter statistics. However, we confirm as it was recently shown in the context of Ly$\alpha$ forest tomography that the Poisson-lognormal model provides the correct two-point statistics or power-spectrum.
Please note, how the power-spectrum (green curve in panel f)  of the converged samples are similar to the underlying power-spectrum (dark blue curve in panel f) and differ from the prior power-spectrum (red curve).

\section{Discussion}

We have presented the Bayesian approach to infer density fields and power-spectra in the context of large-scale structure analysis from non-Gaussian distributed data. Although the results are very encouraging { especially for matter field estimations in high-density regions and power-spectrum estimation} some of the models need to be revised { to get a more detailed characterization of the large-scale structure on small scales}. The lognormal assumption { leads to quantitatively} wrong estimates in under-dense regions at scales below 10 $h^{-1}$ Mpc. At those scales higher order correlation functions start to become relevant. We have shown how this could be modeled with a multivariate Edgeworth expansion. However, the problem of such an approach is that one would need additional models for the higher { order} correlation { functions} introducing hereby more parameters. A different ansatz based on a physical approach would be required to solve this problem. { Focusing on the Gaussian component of the density field and encoding the nonlinear transformation in the likelihood is a very promising approach as it radically simplifies the problem.}
 We have addressed other issues like the non-Poisson character of the galaxy distribution and how this could be implemented in a Bayesian context. 
Although the Bayesian techniques available are powerful enough to deal with complex problems we think that much more effort has to be done in this direction by studying the large-scale structure from simulations and extracting precise statistical  models.

\begin{acknowledgement}
The author thanks the Ludwig Maximilians University, the Max-Planck Institute for Extraterrestrial Physics and the  Max-Planck Institut for Astrophysics for their hospitality and technical support. 
\end{acknowledgement}

{\small
\bibliographystyle{spphys}
\bibliography{lit}
}

\end{document}